# Two Fish Encryption Based Blockchain Technology for Secured Data Storage

Dinesh Kumar K[1], Duraimutharasan N[2]

[1,2]Department of Computer Science and Engineering, AMET University, India

[1]dineshkumar01@gmail.com, [2]duraibose@gmail.com

**ABSTRACT**

*Data security and sharing remains nuisance among many applications like business data, medical data, banking data etc. In this research, block chain technology is built with encryption algorithm for high level data security in cloud storage. Medical data security seems critical aspect due to sensitivity of patient's information. Unauthorized access of medical data creates major issue to patients. This article proposed block chain with hybrid encryption technique for securing medical data stored in block chain model at cloud storage. New Two fish encryption model is implemented based on RSA Multiple Precision Arithmetic (MPA). MPA works by using library concept. The objective of using this methodology is to enhance security performance with less execution time. Patient data is processed by encryption algorithm and stored at blockchain infrastructure using encrypted key. Access permission allows user to read or write the medical data attached in block chain framework. The performance of traditional cryptographic techniques is very less in providing security infrastructure. Proposed blockchain based Two fish encryption technique provides high security in less encryption and decryption time.*

*Keywords: Data security, Blockchain, Two fish encryption, cloud computing, medical data security, RSA-Multiple precision arithmetic.*

## 1. INTRODUCTION

In the modern medical field, medical data has been used for the invention of recent strategies and healing procedures for curing diseases [1]. The medical data is very sensitive aspect where patients do not like to share with others. Security of medical data storage can be ensured by using two techniques. In the first technique, medical information is stored in the database locally and set up a privilege to access the medical information. In the second approach, the stored clinical data encrypted using patient's key value and in future it can be used by the patient's key. The main dilemma of the first approach, locally stored medical data may be modified or deleted. Also, it cannot be shared with doctors. During the diagnosis phase of the disease and treatment were taken by a patient, the key should not be shared with others, it creates a problem with the second approach. Above crisis will damage the availability of medical data in the local storage database. The key force for the above-stated declaration is due to the digitized medical data and accessing it employing professionals is suggested by recent articles [2] [3]. To improve medical information governance and safety regulations like Health Insurance Portability and Accountability Acts (HIPAA) [4] in the USA or the General Data Protection and Regulation (GDPR) [5] at Europe needs high security of sharing the information. Privateness mode of data might cause severe consequences for activities of a healthcare information breach.

The existing cryptographic algorithm in medical data storage methods used a private cloud platform, which carry the limitations on sharing of data and scalability [6]. As blockchain and cloud computing are considered as matured associated strategies which have performed fast development in clinical and fitness services, together with scientific normalization, healthcare services through mobile, e-commerce in medical and on-line mode facility [7]. The block chain system connects the individuals in a P2P form. It includes P2P network design, encryption technology, implementing distributed algorithm and use of data storage [8]. Implementing the limitations of blockchain technology by combining with other cryptographic techniques to discourse the security problems of storing medical data management [9]. Lack of ability and consciousness in implementation, limitations arise in security side of blockchain based cloud storage works quite slow in progress. These above challenges cause delay in the approval of the blockchain technology by the medical institutions. Even though there are numerous start-ups procedures are completely based on blockchain technology, the medical organization refused for using this technology [10]. Our proposed work, hybrid system of two fish with RSA MPA encryption algorithm provides solutions for secure storing of medical data via blockchain based cloud storage in an efficient manner.

In conventional encryption procedure, sender and receiver must generate the public key and private key. Before sending the textual content data, the sender (user) encrypts the textual content using public key of receiver. At the receiving side, the desired client decrypts the textual content data using private key. However, it calls lot of network issues and additionally occupies the memory [11]. But our proposed hybrid system of two fish with RSA MPA encryption algorithm converts the textual content medical data encrypted by medical institution A's public key and decrypted by medical institution B's private key. By this way they can share the medical data [7]. To guarantee the safety and the privateness of medical records, we want to expand a powerful asymmetric cryptographic algorithm is followed to encrypt textual medical data on this work, at a low cost and in an efficient way. A person tries to get medical data, he needs to recognize the corresponding decryption key [12]. Our proposed work provides high dimensionality in the security of sharing medical data among various medical institutions in an efficient way. And also assures keeping the privacy of patient information.

Further, structure of this paper has been planned as follows: Section 2 demonstrate about related survey on block chain security. A proposed methodology structure is explained in Section 3. Section 4 describes about the experimented results and finally Section 5 concludes the paper with future scope.

## 2. RELATED WORKS

Due to the development of medical field, storing of medical data in a secure way considers as a significant role, the conventional centralized scheme of medical data storage has been developed and it does not satisfy the requirements of available data with the high hazard of privacy expose is proposed in paper [13-15]. The various researchers are focused on blockchain technology to provide more secure on medical data. The introduction of blockchain technology creates more efficient infrastructure to manage and maintained the digitalized medical data was suggested by Vazirani et al., [16]. To improve the health care consequences without comprising the safekeeping of patient's information, a feasibility study on utilizing of blockchain technology with cloud storage scheme is developed in paper [17-18].

The combination of blockchain based technology with attribute-based scheme is used to provide security in sharing and storing of medical records and to access digital health care records was suggested in [19]. Another tendency revealed from blockchain technology in the concept of traditional security adopted in a single domain administrative for sharing of medical data is insufficient with multiple healthcare domains. Therefore, advanced cryptographic algorithms are required with the features of rich access control and strict high dimensionality secure enforcement. Nowadays, adopting advanced features of cryptographic algorithms research projects are carried out to provide secure processing of clinical data in the cloud storage [9].

To provide a more efficient and friendly service for medical data storage schemes, various solutions are available at cloud technology. Security management was proposed in paper [20-26]. Patient's medical information is a crucial thing, to store in high secure and privacy with cloud-based storage platform. To guarantee the security and data privacy over a patient's data, we should implement a smart storage method which include the smart IoT-based healthcare architecture is discussed in [27]. Other solutions for sharing delicate medical data on several methods like medical data accumulation of non-standard diagonal method was suggested in [28], sharing of medical data with a cloud-based model used in[29], a hybrid solution of sharing medical data in[30, 31], storage structure of scalable privacy with data preserving scheme[32], a secure system using a fog computation technique in [33], and a distributed based architecture with doubles tag micro aggregation scheme in [34] are implemented. The main problems among these techniques are computational complexity and more time consumption. However, most of the users do not trust the third party of the organization in keeping their medical data secure and in a confidential manner.

Decentralized ledger is used in Blockchain technology to record every medical transaction. It records transaction event as product of source state to present state permanent storage scheme, which was used in paper [35-37]. The features of Blockchain technology are decentralization, immutableness, and verifiability which are essential in the field of medical healthcare, exclusively in the handling of medical records in a secure way.

The improved encrypted version of proxy scheme called Fuzzy based Conditional Identity (FCI), in which exchange of medical data in a privacy-preserving where keys are extracted from user's biometric measures. The content of medical data transactions kept in privacy and consensus efficiency by using blockchain-based medical data storage platform [38]. The ring signature scheme is adopted the elliptic curve model to enhance a privacy medical data storage protocol in user's identity privacy and protection of medical data. The protection of medical data transaction's privacy ring signature scheme is not an applicable one [39]. A new approach of medical data sharing scheme is implemented by combining the ML, Blockchain and cloud storage scheme. This combined scheme can easily and effectively share of medical data transactions between different medical organizations. However, it cannot provide the assurance of receiving exact medical data [40].

By analysing the existing schemes and various traditional methods, it can be found that combining blockchain-based cloud storage in medical institutions has simplified the enrichment of service quality. Preserving of valuable content of medical data is a challenging task between patients and medical research institutions, especially in distributing the data with various entities in smart contracts with all-inclusive privacy considerations. Furthermore, few types of research have focused on this challenging task of whether the collection of medical data obtained from patients that meets their requirements and keep securely is a great challenge [15].

The table 1 shows the related works that can be implemented in security of medical information on blockchain technology.

Table 1: Review on other techniques

| Papers | Description |
| --- | --- |
| Xia et al [41] | smart contracts contain secret keys |
| Omar et al. [42] | Getting decryption key from the owner of medical data. |
| Ferdous et al. [43] | DRAMS is to deploy a decentralized architecture |
| Guo et al. [44] | multi-authority attribute-based signature scheme |
| Zyskind et al [45] | It is a proposed decentralized computation platform |
| Yue et al. [46] | Encrypted data is stored in private blockchain technology with health care data gateway architecture. |
| Alevtina et al. [47] | Encrypted data is stored in cloud sever. |
| Azaria et al. [48] | Accessing rights to get medical data . |

## 3. MEDICAL FILE STORAGE AT BLOCKCHAIN WITH CLOUD STORAGE

In sharing of clinical data/information using blockchain technology, the foremost step is to ensure the reliable in communication and also in security of medical data storage, it is important to build a chain architecture, effectively decide the identification of two entries, that is initiator identity of service and identity of recipient. The system model of medical data storage based on block chain with cloud storage is shown in Fig 1.

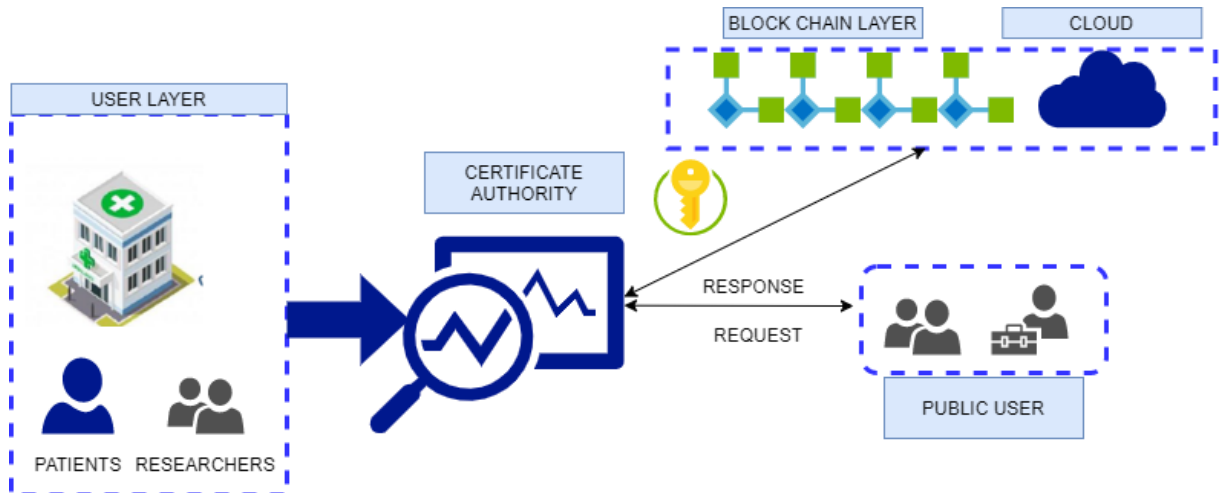

Fig. 1: Medical information storage at block chain with cloud storage

This model is combined into four layers: Certificate Authority, User Layer, Block Chain Layer with cloud storage and public user layer.

**Certificate Authority (CA)**

CA act as a authority provider to generate keys, manage, distribute the digital certificate and system administrator. CA eliminates the malicious nodes and confirm the health of system. For decrypt the data CA uses patient's private key and maintaining perfectness on the information / data which are stored in the block for medical research.

**User Layer**

Different categorize of users are involved in the user layer for their research or other useful purposes in accessing of medical data from the cloud server. All patient's information is maintained as data privacy. The patient gets details of the existing medical records, which are stored in chain or block by using their private key. Example: healthcare organizations like research institutions, medical institutions, research institutions and governmental bodies.

**Block Chain Layer & Cloud storage**

The blockchain layer helps to connect all distributed health field and contracts are responsible for distribution of data across various medical association. In the cloud storage all medical data are collected from patients or from hospitals in different locations or from researchers combined and put into the storage. It is accessed by only authorized users.

**Public User Layer**

It consists of different categorize of users, researchers, general community in medical platforms. They can access the medical information for the need of their medical investigation, gives treatment for the needy people. Through the proper access only we can able to get the storage data for the medical purpose.

**Implementation of Two Fish Algorithm**

Patient's medical records contain diagnosis information, laboratory test reports, medical imaging data like CT, MRI, X-ray image details, treatment details, special examination details are important information. For the development in medical industry, we have to share these medical records among with patients, medical institutions and researchers [49]. This work implements blockchain based cloud storage. Here medical data is divided into multiple encrypted segments or blocks that are interlinked through a hashing function. This paper implements two fish encrypted algorithm. Two fish cryptographic algorithm is a symmetric key block of cipher text with 128 bits block size and the generated key sizes are up to 256 bits. Implementation of two fish encryption algorithm is shown in fig 2.

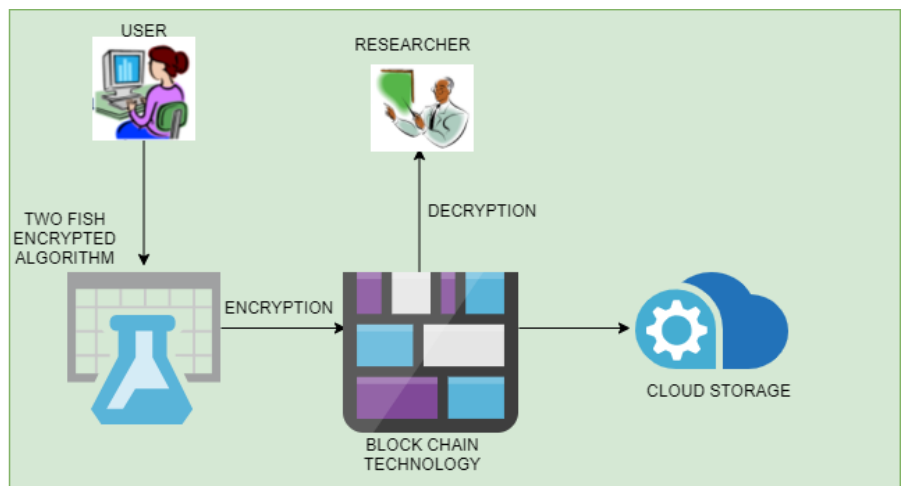

Fig. 2: Implementation of two fish Cryptographic

Implementation steps of Two Fish Cryptographic algorithm is given below [28-30]:

**Algorithm 1:**

**Step 1:** Input block size is 128 bits would be divided into four sections, each for 32 bits words.
**Step 2:** 32-Bit word is XOR input with the four key parts.
$B_{0,i} = R \oplus K_i; \quad i = 0 \text{ to } 3$
Where K is a key and $K_i$ is a sub key i= 0 to 3.
The first key part of word is XOR with $K_0$, second key part of word is XOR with $K_1$ and so on.
**Step 3:** Two fish algorithm uses a Feistel network and it consists of 16 iterations.
**Step 4:** The first key part of word is split up as 4 bytes, where each part is applied to a substituted box. The second key part of word will be first rotated in 8 bits in left and it is also applied to the same set of substitution boxes.
**Step 5:** Diffusing newly substituted data of the 32-bit word, by applying the both the first and second key part of words to MDS matrix (Maximum Distance Separable).
**Step 6:** Then the first key part of word is applied to a pseudo-Hadamard Transform:
pp′ = pp + qq$mmmmmmm$
where p is the first key part of word, q is the second key part of word and p' is the new first key part of word.
**Step 7:** A first key part 'new' is used as input of word p', the second key part of word q is applied to the same transform, which can be represented as: qq′ = pp + 2qq$mmmmmmm$
**Step 8:** Repeat Steps 4 to Steps 7 for 16 iterations.
**Step 9:** The first and second key part of words are swapped with the third and fourth key part of words, the words are XOR to form one more set of round keys for producing the cipher text.

By the above procedure, the medical records are encrypted and stored in blocks. It stores patient's medical report in blockchain and the index value location details are stored in cloud database. Storing of encrypted data and retrieving of decrypted data is done by implementing two fish algorithms [50]. Transaction bodies in medical block chain are Patients, clinical institutions and third-party participants like public users, insurance companies, researchers. Medical records are generated in the medical institutions for the diagnosing the disease and suggest the treatment which is stored at cloud server through block chain. Physicians generates the summaries of medical report of their patients from different medical institutions. These summaries are also processed in the cloud server for storage through block chain. The corresponding patients have ownership for their own medical information. The third-party users or unauthorized users can access this data from chain with proper permission getting by CA. Also, that they provide some services, like recommendation and appointment registration of medical institution. The permissions for the transaction bodies are given in Table 2.

Table 2: Access Permission mode

| Mode of Permission | Patients | Medical Industry | Unauthorized Users |
|---|---|---|---|
| Read access their own medical data | No permission needed | No permission needed | No permission needed |
| Write access their own medical data | No permission needed | No permission needed | No permission needed |
| Read access to third party medical data | Need permission | Need permission, in case of emergency they can access the data | Need permission |
| Write permission to third party clinical data | Need permissions | Need permissions | Need permissions |

The blockchain is responsible for the creation blocks with medical data. When newly medical data is generated for the patient, this is validated and converted into new block, then added to the main chain for the security purpose. The medical data in the blockchain is authenticated by two fish encryptions for the security purpose. For sharing authorization medical data with another medical institution, it needs to get public key of receiver's medical institution. When a user sends a request to access these medical data along with public key, encrypted text of data will return to the user. At the receiving end user decrypts this cipher text files to get the original medical data [51]. If unauthorized user tries to access the medical data, chain cannot allow them to decrypt the medical data. Each medical institution generates a two fish encrypt of medical data D, and encrypts the medical data and stored at cloud at L location using public key value pk of all medical institution to send the medical blockchain.

**3.2 RSA using multiple Precision Arithmetic Library**

For providing security authentication for sharing of medical data, we used block chain of P2P network with all the nodes. Each network node generates two keys; sender encrypts medical data by using public key of receiver. At the same time receiver decrypt the medical data by using private key. For the security authentication scheme, in this work we proposed RSA algorithm using MPA Library (multiple Precision Arithmetic). The implementation steps are given below:

**Algorithm 2:**

**Step 1. Key Generation**
Medical Institution A and Medical Institution B request the CA to generate their public key and private key. Sender encrypts medical data by using public key of receiver. At the same time receiver decrypt the medical data by using private key. For the key generation, RSA algorithm uses two large prime numbers $a, b$.
1. Select $a, b$
2. Calculate $n = a * b$
3. Calculate $\emptyset(n) = (a - 1) * (b - 1)$
4. Select integer $e$ and $e$ is a public key.
5. $gcd(\emptyset(n), e) = 1; 1 < e < \emptyset(n)$
6. Calculate $d$ and $d$ is a private key.
7. $d = e^{-1} mod\ \emptyset(n)$
8. Public key: $pubkey = e, n$
9. Private Key: $prikey = d, n$

Now Public as well as private key for medical institution A & B is generated.

**Step 2: Encryption file generation**
Encrypted file E1 is generated, at first RSA-MPA with public key is used by Medical Institution A. By using second layer, encrypted file E2 is generated by Medical Institution A. Uploading the encrypted files E1 & E2 of Medical Institution A into the server.
$$Encrypt = encr(m, a_k)$$
**Step 3:** Make $m$ is the encrypted information.
**Step 4:** Calculate cipher text by using following formula,
$$C = P^e mod\ n, P < n$$
$C$ is Cipher text; $P$ is Palin text; $e$ is a Encryption key and n is a block size.
**Step 5:** Construct decryption key
$$P = C^d mod\ n$$
Medical Institution A encrypts medical data by using public key of Medical Institution B. At the same time Medical Institution B decrypt the medical data by using private key of Medical Institution A.
**Step 6:** Calculate $Rpubk\ (public\ key)$ and $Rprik$ ( private key)
**Step 7:** The decrypted cipher text with key $Rpubk$ is generated and transmit to the cloud server.
**Step 8:** By using this algorithm, the public key $Rpubk$ is generated with the second layer of cipher text, for construct decryption key with public key $Rpubk$ is generated with the first-layer of cipher text.
**Step 9:** Generate new cipher text, the cloud server uses the decryption key uploaded by Medical Institution A.
**Step 10:** Medical Institution B requests data and $decrpt \rightarrow decrypt(d, b_k)$
**Step 11:** Medical Institution B requests to decrypt the data in cloud. The cloud server sends the decrypted text to Medical Institution B and decryption uses RSA to obtain the original text data [7]. The working principle of this algorithm is shown in the Figure 3.

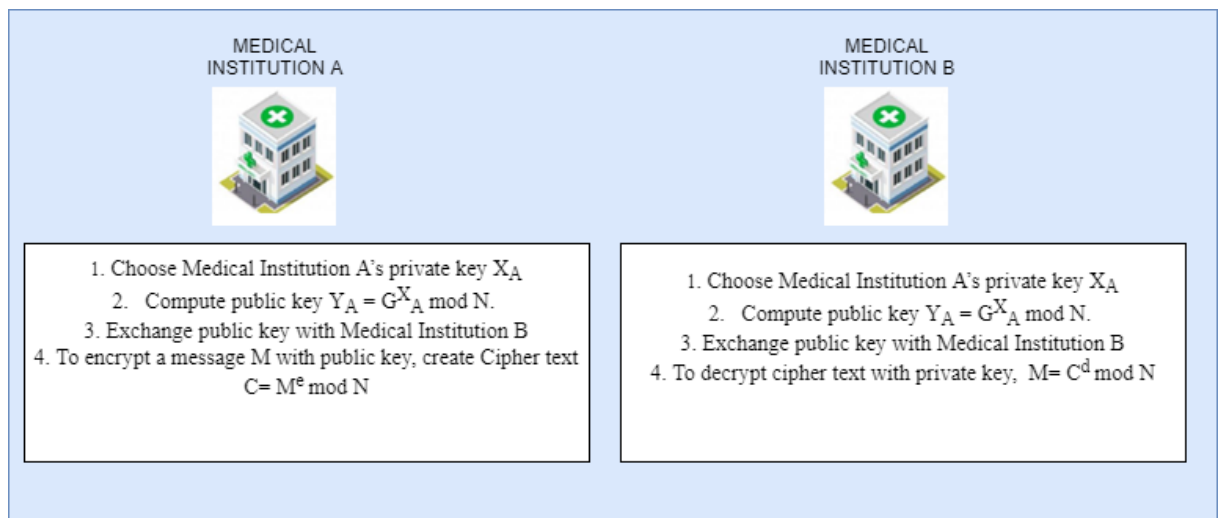

Fig. 3: Working Principle of RSA using MPAL

In this work Public Key is used for encryption of medical data and Private Key is used for decryption of medical data. The MPA library is used to provide the fastest key generation, encryption and decryption routines.

**3.3 Proposed Hybrid system of Two Fish with RSA-MPA Library Encryption Algorithm**

In general, using hybrid method for encryption, ciphers medical data with public and private keys are highly protected while sharing of medical data [52]. Figure 4 shows the proposed (Two Fish + RSA) hybrid architecture. In the beginning key will be generated with two fish algorithm and medical data uses RSA for encryption using MPAL. Finally, encryption of the cipher text medical data is processed at receiver side.

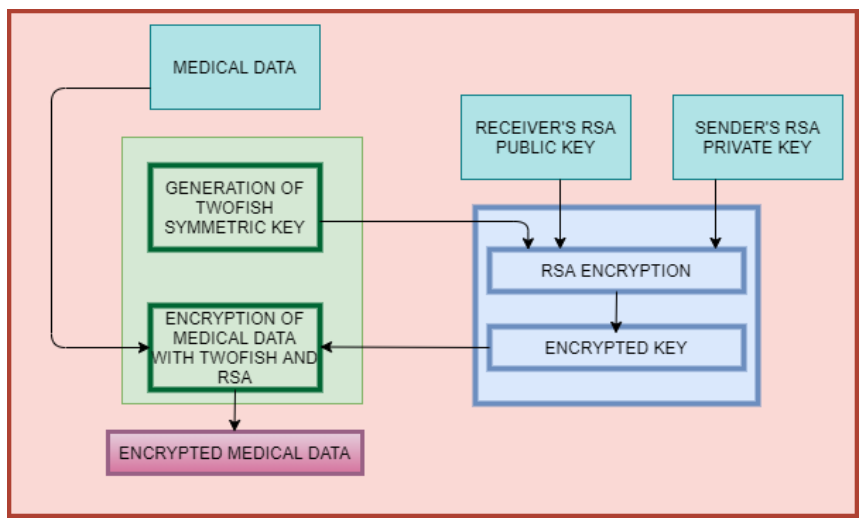

Fig. 4: Proposed (Two Fish + RSA MPA) hybrid architecture.

When compared with other traditional algorithms like key-aggregate cryptosystem(KAC), Attribute-based encryption (ABE) requires public key for encryption and it is fully dependent on attributes [53]. Similarly, compared with other existing algorithms with the concept of privacy protection and secure storage of tamper-proof algorithm will not work effectively. Taking into encryption time of ABE, KAC, two fish, AES algorithm our proposed hybrid algorithm of Two fish and RSA MPA algorithm requires less time and also keeps high security level.

**4. PERFORMANCE ANALYSIS**

The main contribution of our work is to provide security in storing and sharing of medical data. Algorithm 1 describes the design of two fish encrypted concept. It interacts with blockchain based cloud storage, and access the stored medical records with proper permission assign to the user. By using JMeter, accessing of medical data is analysed. Latency is calculated by number of user's requests between 10 and 100 within the time periods of 2, 5, 15, 20 and 45 minutes. The latency time has been calculated by evaluating the time occupied to deliver the data by user's request. This is shown in table 3.

Table 3: Latency time per number of user's requests

| No. of active Users | Latency time (Sec) |
|---|---|
| 10 | 155.67 |
| 20 | 256.78 |
| 30 | 361.34 |
| 40 | 457.56 |
| 50 | 560.13 |
| 60 | 680.89 |
| 70 | 767.12 |
| 80 | 860.67 |
| 90 | 976.45 |
| 100 | 1150.46 |

An important observation of table 3 shows on the latency time is increases as per the request of user increases. This happens because of trade between the securities on medical data over low latency. Even though the latency time increases but efficiency is maintained by two fish encryption algorithm.

The speed of program is evaluated using execution process based upon the encryption and decryption of the file with a different size. Comparison table of encryption and decryption using various methods is given in the table 4.

Table 4: Execution time for Encryption & Decryption in seconds

| File Name | File Size | Two fish [29] | | RSA | | Two fish+RSA[32] | | Proposed Two fish+RSA MPA | |
|---|---|---|---|---|---|---|---|---|---|
| | | Encrypt | Decrypt | Encrypt | Decrypt | Encrypt | Decrypt | Encrypt | Decrypt |
| 100.txt | 100 kb | 0,093 | 0,031 | 0,085 | 0,029 | 0,081 | 0,025 | 0,075 | 0,020 |
| 200.txt | 200 kb | 0,234 | 0,047 | 0,201 | 0,041 | 0,198 | 0,035 | 0,191 | 0,030 |
| 300.txt | 300 kb | 0,312 | 0,078 | 0,298 | 0,071 | 0,289 | 0,067 | 0,250 | 0,057 |
| 400.txt | 400 kb | 0,421 | 0,125 | 0,395 | 0,121 | 0,389 | 0,118 | 0,370 | 0,115 |
| 500.txt | 500 kb | 0,765 | 0,187 | 0,689 | 0,182 | 0,675 | 0,175 | 0,620 | 0,160 |

**Security Analysis**

In accessing medical data in a secure way, we proposed hybrid of two fish and RSA MPA Library access data successfully, must meet some criteria and decrypt the medical data. Table 5 shows the comparison between some traditional methods and the proposed scheme.

Table 5: Comparison between proposed system with other traditional systems.

| Methods | Privacy | Integrity | Anonymous | Attack Resistance | Tamper Proof | Less Computing time |
|---|---|---|---|---|---|---|
| PoW | No | Yes | No | Yes | Yes | No |
| MedShare | No | Yes | Yes | Yes | Yes | No |
| MedBlock | No | Yes | Yes | Yes | Yes | No |
| DACC | No | No | No | Yes | Yes | No |
| PoS | Yes | Yes | No | No | No | Yes |
| Our proposed Scheme | Yes | Yes | Yes | Yes | Yes | Yes |

In the observation of the above table 5, the blockchain system with cloud storage plays a significant role in accessing of clinical data in secure way and sharing. To streamline and perform encryption as well as decryption of any given medical data passed in our proposed scheme of hybrid system of two fish with RSA MPA library measures the better performance on the basis of privacy, integrity, anonymous, attack resistance, tamper proof and less computing time produces more secure in storing and sharing of medical data through blockchain technology. For experimental purposes our proposed work, hybrid system of Two fish with RSA MPA library algorithm stored medical data in various sized text files and the outcomes gives the best result in terms of encrypted and decrypted time of the medical data. Key length uses 256 bits and 16 bytes of block size is used in Two fish algorithm. For the proposed work RSA-MPAL algorithm generates 2048 bits key pair. This technique can encrypt medical data using public key as well as decrypts the data using private keys. The benefits of proposed work are speed of encryption and decryption time because it uses MPA library concept.

The analysis of the result of encrypted data time calculation is shown in Fig. 5

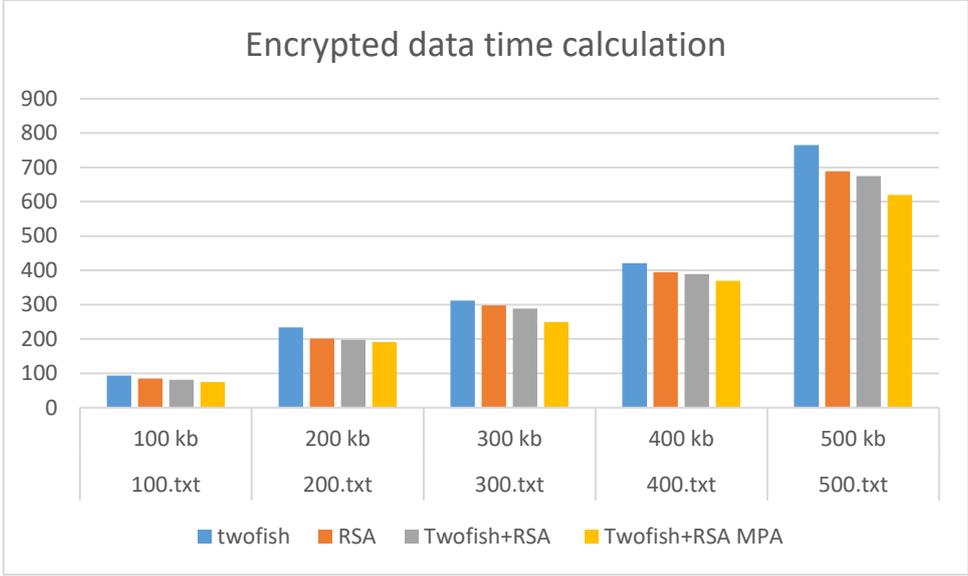

Fig. 5: Analysis of Encrypted Data time calculation in seconds

In the observation, it shows, that throughout the encryption medical data text file, the encryption time is increased proportionally depends upon the size of medical data text file. Comparison results on algorithms Twofish, RSA, Twofish + RSA and Twofish + RSA MPA in terms of encryption time criteria the Twofish + RSA MPA algorithm is better and needs less time comparatively to other algorithms. The analysis of the result of encrypted data time calculation is shown in Figure 6.

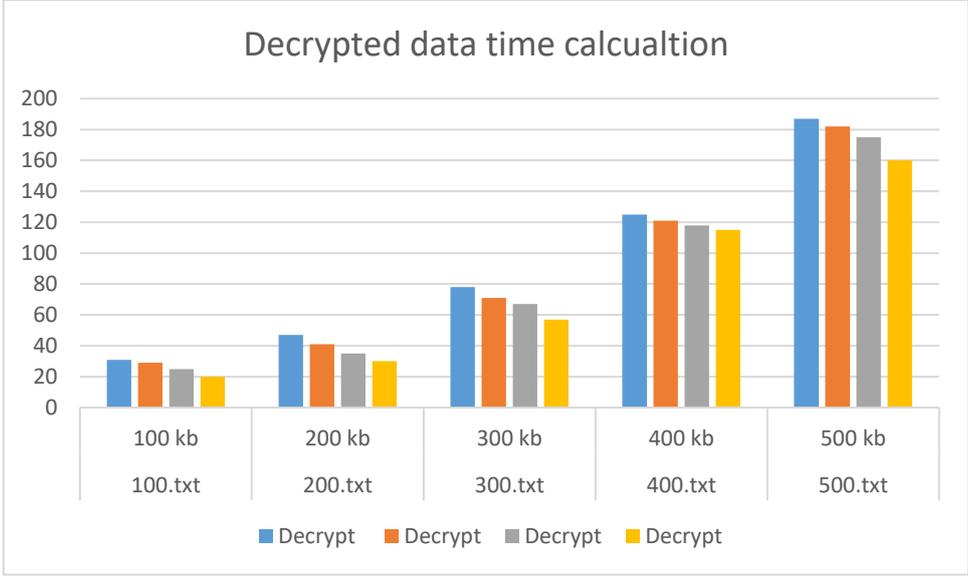

Fig. 6: Analysis of Decrypted Data time calculation in seconds

Comparison results on algorithms Two fish, RSA, two fish + RSA and Two fish + RSA MPA in terms of decryption time criteria the Two fish + RSA MPA algorithm is better and needs less time comparatively to other algorithms.

## 5. CONCLUSION

Hybrid system of Twofish with RSA MPA library algorithm successfully implemented to maintain high secure of medical data sharing through blockchain based cloud storage. This paper presents an analysis of Twofish, RSA, Twofish + RSA and Twofish with RSA-MPA algorithm in terms of size of file, level of security, latency, time taken to encrypt the file, time taken to decrypt the filewere used. This research helps to conclude that among the provided criteria, new hybrid system of Twofish with RSA MPA takes all benefits from traditional methods so it is significantly secure and faster retrieval of medical data. For future work, proposed hybrid models can be implemented by entropy index value. Additionally, our hybrid system could be improved through analysis and implementation of high-performance message passing computing library, such as, Message Passing Interface (MPI).

## REFERENCES

[1] E. R. Weitzman, L. Kaci, and K. D. Mandl, ``Sharing medical data for health research: The early personal health record experience,'' J. Med.Internet Res., vol. 12, no. 2, pp. 1-10, 2010.

[2] D. B. Taichman et al., ``Sharing clinical trial data: A proposal from the international committee of medical journal editors free,'' PLoSMed.,vol. 13, no. 1, pp. 505-506, Apr. 2016.

[3] Q. Xia, E. B. Sifah, K. O. Asamoah, J. Gao, X. Du and M. Guizani, "MeDShare: Trust-Less Medical Data Sharing Among Cloud Service Providers via Blockchain," in IEEE Access, vol. 5, pp. 14757-14767, 2017, doi: 10.1109/ACCESS.2017.2730843.

[4] Summary of the HIPAA Security Rule. [Online] ,2017 Available:https://www.hhs.gov/hipaa/for-professionals/security/laws-regulations/

[5] General Data Protection Regulation. [Online]. Available:https://eugdpr.org/the-regulation/, 2016.

[6] S. Nepal, R. Ranjan, and K.-K. R. Choo, ``Trustworthy processing of healthcare big data in hybrid clouds,'' IEEE Trans. Cloud Computer., vol. 2,no. 2, pp. 7884, Mar./Apr. 2015.

[7] Liang Huang and Hyung-Hyo Lee, "A Medical Data Privacy Protection Scheme Based on Blockchain and Cloud Computing", Wireless Communications and Mobile Computing Volume 2020, Article ID 8859961, 11 pageshttps://doi.org/10.1155/2020/8859961.

[8] Z. Wan, N. Xiong, N. Ghani, A. V. Vasilakos, and L. Zhou,"Adaptive unequal protection for wireless video transmission over IEEE 802.11e networks," Multimedia Tools and Applications, vol. 72, no. 1, pp. 541–571, 2014.

[9] HaoJin , Yan Luo, Peilong Li And Jomol Mathew,"A Review of Secure and Privacy-Preserving Medical Data Sharing, IEEE Access, 2019.

[10]. Goodarzian, F.; Shishebori, D.; Nasseri, H.; Dadvar, F. A bi-objective production-distribution problem in a supply chain network under grey flexible conditions. RAIRO Oper. Res. 2021, 55, 1287.

[11] S. Tanwar, K. Parekh, and R. Evans, "Blockchain-based electronic healthcare record system for healthcare 4.0 applications, "Journal of Information Security and Applications, vol. 50, no. 2, p. 102407, 2020.

[12] Fan, K., Wang, S., Ren, Y. et al. MedBlock: Efficient and Secure Medical Data Sharing Via Blockchain. J Med Syst 42, 136 (2018). https://doi.org/10.1007/s10916-018-0993-7

[13] Zhang Y ,Xu C ,Li H ,Yang K ,Zhou J ,Lin X . Health Dep: an efficient and secure deduplication scheme for cloud-assisted eHealth systems. IEEE Trans. Ind. Inform. 2018.

[14] Miao Y, Tong Q, Choo K-KR, Liu X, Deng RH, Li H. Secure Online/offline data sharing framework for cloud-assisted industrial internet of things. IEEE Internet Things J. [Internet] 2019;6(5):8681–91. Available from: https://ieeexplore.ieee.org/document/8736869/

[15] Liang J ,Qin Z ,Xiao S ,Zhang J ,Yin H ,Li K . Privacy-preserving range query over multi-source electronic health records in public clouds. J. Parallel Distributed Computer. 2020.

[16] Vazirani AA, O'Donoghue O, Brindley D, Meinert E. Blockchain vehicles for efficient medical record management. npj Digit Med [Internet]. 2020;3(1):1–5. Available from: http://dx.doi.org/10.1038/s41746- 019- 0211- 0

[17] Ivan D . Moving toward a blockchain-based method for the secure storage of patient records. NIST Work Blockchain Health. 2016.

[18] Haiping Huang, Peng Zhu,  Fu Xiao , Xiang Sun  , Qinglong Huang," A blockchain-based scheme for privacy-preserving and secure sharing of medical data", 2020 Elsevier Ltd.

[19] R. Guo, H. Shi, Q. Zhao, and D. Zheng, ``Secure attribute-based signature scheme with multiple authorities for blockchain in electronic health records systems,'' IEEE Access, vol. 6, pp. 11676-11686, 2018.

[20] Zhang, Y., Qiu, M., Tsai, C., Hassan, M., and Alamri, A., Health- CPS: Healthcare-CPS: Healthcare Cyber-Physical System Assisted by Cloud and Big Data. IEEE Syst. J. 11:88–95, 2017. https://doi.org/10.1109/JSYST.2015.2460747.

[21] Bahga, A., and Madisetti, V., A Cloud-based Approach for Interoperable Electronic Health Records (EHRs). IEEE J Biomed Health 17:894–906, 2013. https://doi.org/10.1109/JBHI.2013. 2257818.

[22] Godinho, T., Viana-Ferreira, C., and Silva, L., A Routing Mechanism for Cloud Outsourcing of Medical Imaging Repositories. IEEE J Biomed Health 20:367–375, 2016. https:// doi.org/10.1109/JBHI.2014.2361633.

[23] He, C., Fan, X., and Li, Y., Toward Ubiquitous Healthcare Services with a Novel Efficient Cloud Platform. IEEE Bio-med Eng 60:230– 234, 2013. https://doi.org/10.1109/TBME.2012.2222404.

[24] Wang, H., Ding, S., Wu, D., Zhang, Y., and Yang, S., Smart connected electronic gastro scope system for gastric cancer screening using multi-column convolutional neural networks. Int. J. Prod. Res.:1–12, 2018. https://doi.org/10.1080/00207543.2018.1464232.

[25] Ding, S., Li, Y.,Wu, D., Zhang, Y., and Yang, S., Time-aware cloud service recommendation using similarity-enhanced collaborative filtering and ARIMA model. Decis. Support. Syst. 107:103–115, 2018. https://doi.org/10.1016/j.dss.2017.12.012.

[26] Ding, S., Wang, Z., Wu, D., and Olson, D. L., Utilizing customer satisfaction in ranking prediction for personalized cloud service selection. Decision Support. Syst. 93:1–10, 2017. https://doi.org/10. 1016/j.dss.2016.09.001.

[27] Yang, Y., Zheng, X., Guo, W., Liu, X., and Chang, V., Privacy preserving Smart IoT-based Healthcare Big Data Storage and Self-adaptive Access Control System. Inf. Sci.:1–26, 2018. https://doi.org/10.1016/j.ins.2018.02.005.

[28] Singh, K., and Batten, L., Aggregating Privatized Pedical Data for Secure Querying Applications. 72:250– 263, 2017. https://doi.org/10.1016/j.future.2016.11.028.

[29] Alshagathrh, F., Khan, S., Alothmany, N., Al-Rawashdeh, N., and Househ, M., Building a Cloud-based Data Sharing Model for the Saudi National Registry for Implantable Medical Devices: Results of a Readiness Assessment. Int. J. Med. Inform. 118:113–119, 2018. https://doi.org/10.1016/j.ijmedinf.2018.08.005.

[30] Yang, J., Li, J., and Niu, Y., A Hybrid Solution for Privacy Preserving Medical Data Sharing in the Cloud Environment. Future General  43-44:74–86, 2015. https://doi.org/10. 1016/j.future.2014.06.004.

[31] Xia, C., Meloni, S., Perc, M., and Moreno, Y., Dynamic instability of cooperation due to diverse activity patterns in evolutionary social dilemmas. EPL 109:58002, 2015. https://doi.org/10.1209/0295-5075/109/58002.

[32] Jabeen, F., Hamid, Z., and Abdul, W., Enhanced Architecture for Privacy Preserving Data Integration in a Medical Research Environment. IEEE Access 5:13308–13326, 2017. https://doi.org/10.1109/ACCESS.2017.2707584.

[33] Al, H. H., Rahman, S., and Hossain, M., A Security Model for Preserving the Privacy of Medical Big Data in a Healthcare Cloud Using a Fog Computing Facility with Pairing-based Cryptography. IEEE Access 5:22313–22328, 2017. https://doi.org/10.1109/ ACCESS.2017.2757844.

[34] Solanas, A., Martínez-Ballesté, A., and Mateo-Sanz, J., Distributed Architecture with Double-phase Microaggregation for the Private Sharing of Biomedical Data in Mobile Health. IEEE 8:901–910, 2013. https://doi.org/10.1109/TIFS.2013.2248728.

[35] Griggs, K. N., Ossipova, O., Kohlios, C. P., Baccarini, A. N., Howson, E. A., and Hayajneh, T., Healthcare Blockchain System Using Smart Contracts for Secure Automated Remote Patient Monitoring. J. Med. Syst. 42:130, 2018. https://doi.org/10.1007/s10916-018-0982-x.

[36] Lin, C., He, D., Huang, X., Choo, K.-K. R., and Vasilakos, A. V., BSeIn: A blockchain-based secure mutual authentication with fine-grained access control system for industry 4.0. J. Network. Computer Appl. 116:42–52, 2018. https://doi.org/10.1016/J. JNCA.2018.05.005.

[37] Lin, C., He, D., Huang, X., Khan, M., and Choo, K., A New Transitively Closed Undirected Graph Authentication Scheme for Blockchain-Based Identity Management Systems. IEEE Access 6: 28203–28212, 2018. https://doi.org/10.1109/ACCESS.2018. 2837650.

[38] Fimiani G. Supporting privacy in a cloud-based health information system by means of fuzzy conditional identity-based proxy re-encryption (FCI-PRE). In: Proceedings - 32nd IEEE International Conference on Advanced Information Networking and Applications Workshops, WAINA 2018. 2018.

[39] Li X ,Mei Y . A Blockchain Privacy Protection Scheme Based on Ring Signature. IEEE Access 2020; 8:76765–72 . 2020

[40] Zheng X ,Mukkamala RR ,Vatrapu R ,Ordieres-Mere J .," Blockchain-based personal health data sharing system using cloud storage", In: 2018 IEEE 20th International Conference on e-Health Networking, Applications and Services, Healthcom 2018.

[41] Xia, Q., Sifah, E. B., Asamoah, K. O., Gao, J., Du, X., and Guizani, M., MeD Share: trust less medical data sharing among cloud service providers via blockchain. IEEE Access PP(99):1–1,2017.

[42] Omar, A. A., Rahman, M. S., Basu, A., and Kiyomoto, S.,MediBchain: a blockchain based privacy preserving platform for Healthcare Data. In: International Conference on Security, Privacy and Anonymity in Computation, Communication and Storage, pp. 534–543, 2017.

[43] S. Ferdous, A. Margheri, F. Paci, and V. Sassone, ``Decentralised runtime monitoring for access control systems in cloud federations,'' in Proc. IEEE Int. Conf. Distributed Computer, Jun. 2017, pp. 1-11.

[44] R. Guo, H. Shi, Q. Zhao, and D. Zheng, ``Secure attribute-based signature scheme with multiple authorities for blockchain in electronic health records systems,'' IEEE Access, vol. 6, pp. 11676-11686, 2018.

[45] G. Zyskind, O. Nathan, and A. Pentland. (2015). ``Enigma: Decentralized computation platform with guaranteed privacy.'' [Online]. Available: https://arxiv.org/abs/1506.03471

[46] Yue, X., Wang, H., Jin, D., Li, M., and Jiang, W., Healthcare data gateways: found healthcare intelligence on blockchain with novel privacy risk control. J. Med. Syst. 40(10):1–8, 2016.

[47] Dubovitskaya, A., Xu, Z., Ryu, S., Schumacher, M., and Wang,F., Secure and trustable electronic medical records sharing using Blockchain, preprint arXiv:1709.06528, 2017.

[48] Azaria, A., Ekblaw, A., Vieira, T., and Lippman, A., Med Rec:using blockchain for medical data access and permission management. In: International Conference on Open and Big Data,pp. 25–30, 2016.

[49] Yang, J. J., Li, J. Q., and Niu, Y., A hybrid solution for privacypreserving medical data sharing in the cloud environment. FutureGener. Computer. Syst. 43-44(45):74–86, 2015.

[50] Xu Cheng&Fulong Chen & Dong Xie& Hui Sun& Cheng Huang," Design of a Secure Medical Data Sharing SchemeBased on Blockchain", Journal of Medical Systems (2020) 44:52https://doi.org/10.1007/s10916-019-1468-1

[51] Z. Liu, B. Hu, B. Huang, L. Lang, H. Guo, and Y. Zhao, "Strategies of Haze Risk Reduction Using the Tripartite Game Model," Complexity, vol. 2020, Article ID 2145951, 11 pages, 2020.

[52] E. Jintcharadze and M. Iavich, "Hybrid Implementation of Twofish, AES, ElGamal and RSA Cryptosystems," 2020 IEEE East-West Design & Test Symposium (EWDTS), 2020, pp. 1-5, doi: 10.1109/EWDTS50664.2020.9224901

[53] Yi Chen & Shuai Ding & Zheng Xu &Handong Zheng & Shanlin Yang," Blockchain-Based Medical Records Secure Storage and Medical Service Framework", Journal of Medical Systems (2019) 43: 5https://doi.org/10.1007/s10916-018-1121-4